\input epsf.tex
\documentclass[12pt]{article}
\usepackage{graphicx}
\usepackage{psfig,picinpar,floatflt,amssymb,epsf}
\csname @addtoreset\endcsname{equation}{section}

\def\bseq{\begin{subequation}}  
\def\eseq{\end{subequation}}
\def\bsea{\begin{subeqnarray}}  
\def\esea{\end{subeqnarray}}

\newcommand{\beq}{\begin{equation}}
\newcommand{\eeq}{\end{equation}}
\newcommand{\bea}{\begin{eqnarray}}
\newcommand{\eea}{\end{eqnarray}}
\newcommand{\ena}{\end{eqnarray}}

\renewcommand{\a}{\alpha}
\renewcommand{\b}{\beta}

\renewcommand{\d}{\delta}
\newcommand{\pa}{\partial}
\newcommand{\g}{\gamma}
\newcommand{\G}{\Gamma}

\newcommand{\D}{\Delta}
\newcommand{\e}{\epsilon}

\renewcommand{\L}{\Lambda}

\newcommand{\p}{\pi}

\newcommand{\Phib}{\overline{\Phi}}
\newcommand{\adot}{\dot{\alpha}}
\newcommand{\bdot}{\dot{\beta}}
\newcommand{\gdot}{\dot{\gamma}}

\def\Mb{\kern 2pt\mathchoice
        {
         \vbox{\hrule width10pt height 0.4pt depth 0pt
         \kern 1.2pt\hbox{\kern -2pt$\displaystyle M$}}}
        {
         \vbox{\hrule width10pt height 0.4pt depth 0pt
         \kern 1.2pt\hbox{\kern -2pt$\textstyle M$}}}
        {
\vbox{\hrule width6pt height 0.4pt depth 0pt
         \kern 1.0pt\hbox{\kern -2pt$\scriptstyle M$}}}
        {
         \vbox{\hrule width5pt height 0.4pt depth 0pt
         \kern 0.8pt\hbox{\kern -2pt$\scriptscriptstyle M$}}}}

\def\Sb{\kern 2pt\mathchoice
        {
         \vbox{\hrule width6pt height 0.4pt depth 0pt
         \kern 1.2pt\hbox{\kern -2pt$\displaystyle S$}}}
        {
         \vbox{\hrule width6pt height 0.4pt depth 0pt
         \kern 1.2pt\hbox{\kern -2pt$\textstyle S$}}}
        {
         \vbox{\hrule width3.5pt height 0.4pt depth 0pt
         \kern 1.0pt\hbox{\kern -2pt$\scriptstyle S$}}}
        {
         \vbox{\hrule width3pt height 0.4pt depth 0pt
         \kern 0.8pt\hbox{\kern -2pt$\scriptscriptstyle S$}}}}

\def\Rb{\kern 2pt\mathchoice
        {
         \vbox{\hrule width5.5pt height 0.4pt depth 0pt
         \kern 1.2pt\hbox{\kern -2.5pt$\displaystyle R$}}}
        {
         \vbox{\hrule width5.5pt height 0.4pt depth 0pt
         \kern 1.2pt\hbox{\kern -2.5pt$\textstyle R$}}}
        {
         \vbox{\hrule width3.5pt height 0.4pt depth 0pt
         \kern 1.0pt\hbox{\kern -2.2pt$\scriptstyle R$}}}
        {
         \vbox{\hrule width3pt height 0.4pt depth 0pt
         \kern 0.8pt\hbox{\kern -2.2pt$\scriptscriptstyle R$}}}}

  \def\pp{{\mathchoice
      {
          \kern 1pt%
          \raise 1pt
          \vbox{\hrule width5pt height0.4pt depth0pt
            \kern -2pt
            \hbox{\kern 2.3pt
              \vrule width0.4pt height6pt depth0pt
              }
            \kern -2pt
            \hrule width5pt height0.4pt depth0pt}%
            \kern 1pt
       }
        {
          \kern 1pt%
          \raise 1pt
          \vbox{\hrule width4.3pt height0.4pt depth0pt
            \kern -1.8pt
            \hbox{\kern 1.95pt
              \vrule width0.4pt height5.4pt depth0pt
              }
            \kern -1.8pt
            \hrule width4.3pt height0.4pt depth0pt}%
            \kern 1pt
        }
        {
          \kern 0.5pt%
          \raise 1pt
          \vbox{\hrule width4.0pt height0.3pt depth0pt
            \kern -1.9pt  
            \hbox{\kern 1.85pt
              \vrule width0.3pt height5.7pt depth0pt
              }
            \kern -1.9pt
            \hrule width4.0pt height0.3pt depth0pt}%
            \kern 0.5pt
        }
        {
          \kern 0.5pt%
          \raise 1pt
          \vbox{\hrule width3.6pt height0.3pt depth0pt
            \kern -1.5pt
            \hbox{\kern 1.65pt
              \vrule width0.3pt height4.5pt depth0pt
              }
            \kern -1.5pt
            \hrule width3.6pt height0.3pt depth0pt}%
            \kern 0.5pt
        }
    }}

  \def\mm{{\mathchoice
               {
                 \kern 1pt
           \raise 1pt    \vbox{\hrule width5pt height0.4pt depth0pt
                  \kern 2pt
                  \hrule width5pt height0.4pt depth0pt}
                 \kern 1pt}
               {
                \kern 1pt
           \raise 1pt \vbox{\hrule width4.3pt height0.4pt depth0pt
                  \kern 1.8pt
                  \hrule width4.3pt height0.4pt depth0pt}
                 \kern 1pt}
               {
                \kern 0.5pt
           \raise 1pt
                \vbox{\hrule width4.0pt height0.3pt depth0pt
                  \kern 1.9pt
                  \hrule width4.0pt height0.3pt depth0pt}
                \kern 1pt}
               {
               \kern 0.5pt
         \raise 1pt  \vbox{\hrule width3.6pt height0.3pt depth0pt
                  \kern 1.5pt
                  \hrule width3.6pt height0.3pt depth0pt}
               \kern 0.5pt}
               }}

\def\pd{{\kern0.5pt
           + \kern-5.05pt \raise5.8pt\hbox{$\textstyle.$}\kern
0.5pt}}

\def\pmd{{\kern0.5pt
          \pm \kern-5.05pt
\raise6.3pt\hbox{$\textstyle.$}\kern1.5pt}}

\def\md{{\mathchoice
   {
      {{\kern 1pt - \kern-6.2pt \raise5pt\hbox{$\textstyle.$}\kern
1pt}}}
    {
      {{\kern 1pt - \kern-6.2pt \raise5pt\hbox{$\textstyle.$}\kern
1pt}}}
    {
      {\kern0.5pt - \kern-5.05pt
\raise3.4pt\hbox{$\textstyle.$}\kern0.5pt}}
    {
      {\kern0.5pt - \kern-5.05pt
\raise3.4pt\hbox{$\textstyle.$}\kern0.5pt}}}}

\newcommand{\ad}{{\dot{\alpha}}}

\newcommand{\Del}{\nabla}

\newcommand{\boldnabla}{  \nabla \hspace{-0.12in}{\nabla}}

\renewcommand{\th}{\theta}
\renewcommand{\adot}{\dot{\alpha}}
\renewcommand{\bdot}{\dot{\beta}}
\renewcommand{\gdot}{\dot{\gamma}}

\begin{document}

\begin{titlepage}
{\hbox to\hsize{December 2004 \hfill
{Bicocca--FT--04--17}}}

\begin{center}
\vglue .06in
{\Large\bf Covariant quantization of $N=\frac12$ SYM theories\\ [0.3cm]
and supergauge invariance}
\\[.45in]
Silvia Penati\footnote{silvia.penati@mib.infn.it} ~and~
Alberto Romagnoni\footnote{alberto.romagnoni@mib.infn.it}\\
{\it Dipartimento di Fisica dell'Universit\`a degli studi di
Milano-Bicocca,\\ 
and INFN, Sezione di Milano, piazza della Scienza 3, I-20126 Milano,
Italy}\\[.8in]

{\bf ABSTRACT}\\[.0015in]
\end{center}

So far, quantum properties of $N=1/2$ nonanticommutative (NAC) 
super Yang--Mills theories have been investigated in the WZ gauge. 
The gauge independence
of the results requires assuming that at the quantum level
supergauge invariance is not broken by nonanticommutative geometry.
In this paper we use an alternative
approach which allows  studying these theories in a manifestly gauge
independent superspace setup. This is accomplished by generalizing the
background field method to the NAC case, by moving to a momentum
superspace where star products are treated as exponential factors and
by developing momentum supergraph techniques. We compute the one--loop 
gauge effective action for NAC $U({\cal N})$ gauge theories with matter 
in the adjoint
representation. Despite the appearance of divergent contributions
which break (super)gauge invariance, we prove that the effective
action at this order is indeed invariant.


\end{titlepage}


\section{Introduction}

Euclidean superspace geometries with nonanticommutative (NAC) spinorial
variables \cite{ferrara, KPT} have been shown to arise
as field theory limits of strings in flat spacetime when a self--dual
graviphoton background is turned on \cite{peter, OV, seiberg, seiberg2}.

Field theories defined on such geometries describe the
dynamics of bosonic and fermionic degrees of freedom which
realize $(\frac12, 0)$ supersymmetry \cite{seiberg}.
They can be obtained by starting with ordinary extended $N= (1,1)$
supersymmetry and explicitly breaking  it, in a variety of ways,
by turning on nontrivial anticommutation relations for spinorial
coordinates (while in general supersymmetry is broken to
$N=(\frac12, 0)$, there are particular cases where $N=(\frac12, 1)$
supersymmetry survives \cite{ILZ}).

Since in general supersymmetry plays an important role in guaranteing 
renormalization, in the presence of hard breaking it becomes
of primary interest to investigate possible consequences for
renormalizability. 
A first set of results in this direction has been obtained for
the $N=(\frac12, 0)$ WZ model in four dimensions. In particular,
it has been shown \cite{us,us-proc,BF,R,BF2,terashima} that the model,
initially defined by the ordinary WZ action where the products
have been promoted to NAC star products \cite{seiberg} is renormalizable
if extra $F$ and $F^2$ interaction terms are added to the initial
lagrangian, where $F$ is the auxiliary field of the scalar multiplet. 

The most interesting NAC theories to be investigated are supersymmetric
$N=1/2$ gauge theories. They are defined by the ordinary gauge actions
where the products have been promoted to star products.
$U({\cal N})$, $N=1/2$ super Yang--Mills
is the field theory living on a set of ${\cal N}$ coincident D3--branes
of a type IIB superstring
theory compactified on a CY space, in the presence of a 4d self--dual
graviphoton background \cite{OV, seiberg2}.

A general discussion of renormalizability for $N= 1/2$
super Yang--Mills theories based on dimensional analysis
has been given \cite{LR} in the WZ gauge,
while one--loop checks have been performed in \cite{AGS, JONES} 
still in components, in the WZ gauge. An advantage of working in the WZ
gauge is that the expansion of the
gauge superfield strengths in components gives rise to a finite number of
terms also in the NAC case \cite{seiberg, AIO}. In the ordinary
anticommutative case, in the absence of anomalies, renormalizability
in the WZ gauge guarantees renormalizability in any gauge since (super)gauge
invariance is not spoiled by quantum corrections.
However, in the presence of nonanticommutativity we do not have any
{\em a priori} argument to guarantee that (super)gauge invariance will survive
at the quantum level since supersymmetry breaking is realized by
a star product which is defined in terms of ordinary non(super)covariant 
derivatives.
\footnote{An analogous situation is present in 
super Yang--Mills theories defined on noncommutative  
geometries $[x^\mu, x^\nu ] \neq 0$. There, it is well known
\cite{LM, zanon} that off--shell corrections
to the effective action are not gauge invariant term by term, even if
gauge invariance of the complete effective action does not seem to be
affected \cite{PSZ}.}.
Therefore, quantum properties, renormalizability included, of
NAC super Yang--Mills proved in the WZ gauge cannot be safely extended
to any gauge without a deeper understanding of the relation among
nonanticommutativity, susy breaking and (super)gauge invariance at the
quantum level.

To investigate this subject we have been led to develop a different approach to
the $N=1/2$ super Yang--Mills theories which allows for a perturbative
analysis directly in superspace without expanding in components in a
particular gauge.
This has also the advantage of allowing  higher--loop calculations
which in components are usually prohibited by technical difficulties.
We work in $N=(\frac12,0)$ superspace keeping the star product implicit and
performing Fourier transforms  (FT) both in the bosonic and fermionic 
coordinates \footnote{Similar techniques have been used in 
\cite{terashima,BFR3}.}. 
Under FT the star product is traded for exponential factors dependent
on the spinorial momenta and the nonanticommuting matrix. As in the bosonic
case, the diagrams can be classified into ``planar'' and ``nonplanar'',
the nonplanar ones being the only diagrams having a nontrivial exponential
factor dependent on internal and/or external spinorial momenta.
In the ordinary case the supergraph techniques, in
particular the $D$--algebra \cite{superspace} which allows one to reduce
supergraphs to ordinary loop momentum integrals, are easily translated in
momentum superspace as a set of rules for the Fourier tranformed covariant
derivatives $\widetilde{D}$. In the nonanticommuting case, instead,
the presence
of extra exponential factors from the star products affects in a nontrivial
way the spinorial structure of the diagrams and, consequently, the
$\widetilde{D}$--algebra. This is very different from the case
of SYM theories defined on noncommuting superspaces where the
noncommutation of the bosonic coordinates does not modify
the spinorial nature of the supergraphs and $D$--algebra
(or $\widetilde{D}$--algebra) can be performed with the standard rules
of the ordinary commutative case \cite{zanon}.

In the $N=(\frac12, 0)$ superspace setting we use the general procedure
described above to study quantum properties of $U({\cal N})$
super Yang--Mills theories
with (anti)chiral matter in the adjoint representation of the gauge group
\footnote{We note that in the 
presence of star product the group $SU({\cal N})$ is not closed and we are 
forced to consider $U({\cal N})$.}.
In order to carry out the calculations by dealing efficiently with
the classical gauge covariance we generalize the background field method
\cite{GSZ,superspace}
to the nonanticommutative case. The generalization is not
straightforward, primarily due to two reasons: The change of the hermitian conjugation rules in the classical action (the NAC superspace has euclidean signature \cite{KPT, seiberg}) and the lack of basic identities involving covariant
derivatives which in the NAC case are spoiled by the noncommutativity of
the star product.
However, we show that a modified version of the method
exists which allows for a manifestly covariant quantization of gauge
theories, at least at one--loop.

Armed with these techniques we compute one--loop divergent contributions 
to the gauge effective action.
In the presence of spinorial phases coming from the star products the 
divergent nature of the diagrams changes in a nontrivial way. As a
consequence, we find divergent contributions to the two, three and four--point
functions, in contradistinction to the ordinary case where, in the 
background field method approach, only the two--point function is 
divergent \cite{GSZ, superspace}.

The new divergent terms, proportional to the NAC parameter,  
are {\em not} (super)gauge invariant on their own. However, we prove that 
at the level of effective action highly nontrivial cancellations among 
nonvanishing gauge variations occur leading to a one--loop effective
action for the gauge fields which is {\em supergauge invariant}. 
Therefore, supergauge invariance seems to be maintained at the quantum level
despite the presence of the star product which is not manifestly supergauge
invariant. A similar discussion can be found in a paper \cite{JONES} which 
appeared few days ago, when our work was almost finished. There the
authors show that, working in components in the WZ gauge, a one--loop
divergent field redefinition of the gaugino field is needed in order
to restore gauge invariance at one--loop. Our results prove that in
the gauge of superspace no field redefinition is required and the 
effective action turns out to be not only gauge, but even supergauge 
invariant at one--loop.

The paper is organized as follows: In Section 2 we briefly review the
$N=(\frac12,0)$ NAC superspace \cite{seiberg} and supersymmetric
gauge theories defined on it, and formulate the
background field method in the presence of nonaticommutativity.
The Fourier transform to momentum superspace and the corresponding
supergraph techniques are then described in Section 4.
In Section 5 we concentrate on one--loop calculations of two, three and
four--point functions with external vector lines for the $U({\cal N})$
SYM with matter in the adjoint. 
In Section 6 we collect all the results and prove the supergauge 
invariance of the one--loop divergent part of the effective action. 
The last Section is then devoted to some conclusions. 
In this paper we basically list the main results referring to a future 
publication \cite{GPR2} for details and an extended analysis.


\section{SYM theories in $N=1/2$ superspace and background field method}

Nonanticommutative $N=(\frac12 , 0)$ superspace can be defined as
a truncation of euclidean $N=(1,1)$ superspace endowed with nonstandard
hermitian conjugation
rules on the spinorial variables \cite{us, seiberg, ILZ}. It is
described by the set of coordinates $(x^{\a\adot}, \theta^\a ,
\overline{\theta}^{\adot})$, $(\theta^\a)^\dag = i \theta_\a$,
$(\overline{\theta}^{\adot})^\dag = i \overline{\theta}_{\adot}$, satisfying
\beq
\big\{ \theta^{\alpha}, \theta^{\beta} \big\} = 2 {\cal F}^{\alpha \beta}
\qquad \quad {\rm the~ rest} = 0
\label{nc}
\eeq
where ${\cal F}^{\alpha \beta}$ is a $2 \times 2$ symmetric, constant matrix.
This algebra is consistent only in euclidean signature where the chiral
and antichiral sectors are totally independent and not related by
complex conjugation. The euclidean Lorentz group $SO(4) = SU_L(2) \times
SU_R(2)$ is broken to $SU_R(2)$ by (\ref{nc}).

We work in chiral representation \cite{superspace} for supercharges and
covariant derivatives
\bea
&& \overline{Q}_{\adot} = i ( \overline{\pa}_{\adot} - i \theta^{\a} \pa_{\a \adot} )
\quad , \quad Q_{\a} = i \pa_{\a}
\nonumber \\
&& \overline{D}_{\adot} =  \overline{\pa}_{\adot} \qquad \qquad \qquad , \quad
D_{\a} = \pa_{\a} + i \overline{\theta}^{\adot} \pa_{\a \adot}
\label{DQ}
\eea
While the algebra of covariant derivatives is not modified,
the algebra of supercharges receives an extra contribution from (\ref{nc})
and the supersymmetry in the antichiral sector is explicitly broken
\cite{seiberg}.

The NAC geometry (\ref{nc}) can be realized on the class of
smooth superfunctions of $(x^{\a\adot}, \theta^\a ,
\overline{\theta}^{\adot})$, by introducing the nonanticommutative
(but associative) star product
\bea
\phi \ast \psi &\equiv& \phi e^{- \overleftarrow{\pa}_\a {\cal F}^{\a \b}
\overrightarrow{\pa}_\b} \psi
\nonumber \\
&=& \phi \psi - \phi \overleftarrow{\pa}_\a {\cal F}^{\a \b}
\overrightarrow{\pa}_\b \psi
+ \frac12 \phi \overleftarrow{\pa}_\a \overleftarrow{\pa}_\g {\cal F}^{\a \b}
{\cal F}^{\g \d}
\overrightarrow{\pa}_\d \overrightarrow{\pa}_\b \psi
\nonumber \\
&=& \phi \psi - \phi \overleftarrow{\pa}_\a {\cal F}^{\a \b}
\overrightarrow{\pa}_\b
\psi - \frac12 {\cal F}^2 \pa^2\phi \, {\pa}^2 \psi
\label{star}
\eea
where we have defined ${\cal F}^2 \equiv  {\cal F}^{\a \b} {\cal F}_{\a \b}$.
The covariant derivatives (\ref{DQ}) are still derivations for this
product. Therefore, the class of (anti)chiral superfields defined
by the constraints $\overline{D}_{\adot} \ast \Phi = D_\a \ast \overline{\Phi} =0$
are closed.

\vskip 10pt
We now define supersymmetric gauge theories on $N=1/2$ superspace. Since
in the presence of non(anti)commutativity also the $U_\ast(1)$ gauge theory
becomes nonabelian the relations we are going to introduce hold nontrivially
for any gauge group, $U_\ast(1)$ included.

Gauge fields and field strengths together with their superpartners can be
organized into superfields, all expressed in terms of a prepotential
$V$ which is a scalar
superfield in the adjoint representation of the gauge group ($V \equiv
V_a T^a$, $T^a$ being the group generators).

The supergauge transformations are given in terms of two independent
chiral and  antichiral parameter superfields $\L, \overline{\L}$
\beq
e_\ast^V \rightarrow  e_\ast^{V'} = e_\ast^{i \overline{\L}} \ast
e_\ast^V \ast e_\ast^{-i\L}
\eeq
The corresponding covariant derivatives (in gauge chiral
representation) are given by
\beq
 \nabla_A \equiv (\nabla_\a , \nabla_{\adot}, \nabla_{\a \adot})
~=~ ( e_\ast^{-V} \ast D_\a \, e_\ast^V ~,~ \overline{D}_{\adot} ~,~
-i \{ \nabla_\a, \nabla_{\adot} \}_{\ast} )
\eeq
whereas in gauge antichiral representation they are defined as
\beq
 \overline{\nabla}_A \equiv (\overline{\nabla}_\a , \overline{\nabla}_{\adot},
\overline{\nabla}_{\a \adot})
~=~ (D_\a ~,~ e_\ast^{V} \ast \overline{D}_{\adot} \, e_\ast^{-V} ~,~
-i \{ \overline{\nabla}_\a, \overline{\nabla}_{\adot} \}_{\ast} )
\eeq
satisfying $\overline{\nabla}_A = e_\ast^{V} \ast \nabla_A \ast e_\ast^{-V}$.

They can be expressed in terms of ordinary supercovariant derivatives
$D_A, \overline{D}_A$ and a set of connections, as $\nabla_A \equiv D_A - i \G_A$
or $\overline{\nabla}_A \equiv \overline{D}_A - i \overline{\G}_A$.
Nontrivial connections are then
\beq
 \G_\a = ie_\ast^{-V} \ast D_\a \, e_\ast^V
\qquad , \qquad \G_{\a \adot} = -i \overline{D}_{\adot}  \G_\a \label{connections}
\eeq
or
\beq
\overline{\G}_{\adot} = ie_\ast^{V} \ast \overline{D}_{\adot} \, e_\ast^{-V}
\qquad , \qquad \overline{\G}_{\a \adot} = -i D_\a  \overline{\G}_{\adot}
\eeq
The field strengths are defined as $\ast$--commutators of supergauge
covariant derivatives
\beq
 W_\a = -\frac12 [ \nabla^{\adot}, \nabla_{\a \adot} ]_\ast
\qquad , \qquad W_{\adot} = -\frac12 [ \nabla^{\a}, \nabla_{\a \adot} ]_\ast
\eeq
or
\beq
\overline{W}_\a = -\frac12 [ \overline{\nabla}^{\adot}, \overline{\nabla}_{\a \adot} ]_\ast
\qquad , \qquad \overline{W}_{\adot} = -\frac12 [ \overline{\nabla}^{\a},
\overline{\nabla}_{\a \adot} ]_\ast
\eeq
and satisfy the Bianchi's identities $\nabla^\a \ast W_\a + \nabla^{\adot}
\ast W_{\adot} =0$
or $\overline{\nabla}^\a \ast \overline{W}_\a + \overline{\nabla}^{\adot} \ast \overline{W}_{\adot}=0$.

In the presence of chiral matter in the adjoint representation of the
gauge group ${\cal G}$ the SYM action is
\bea
S &=& \int d^4x d^4\th~ {\rm Tr}(e_\ast^{-V} \ast \overline{\Phi} \ast e_\ast^{V} \ast \Phi)
+ \frac{1}{2 g^2} \int d^4x d^2\th ~{\rm Tr}(W^\a W_\a) \nonumber \\
&& - \frac{1}{2} m \int d^4x d^2\th ~\Phi^2 - \frac{1}{2} \overline{m} \int d^4x d^2\overline{\th}~ \overline{\Phi}^2
\label{action}
\eea
where all the superfield can be consistently taken to be real (we are in Euclidean superspace). In what follows we work with ${\cal G} = U({\cal N})$.  
\vskip 10pt
We now generalize the background field method \cite{GSZ, superspace}
to the case of NAC super Yang--Mills
theories with chiral matter in a real representation of the gauge group.
We perform the nonlinear splitting $e_{\ast}^V \rightarrow e_{\ast}^\Omega
\ast e_{\ast}^V$ where $\Omega$ is the background prepotential, and write
the covariant derivatives (in gauge-chiral representation) as

\beq
\nabla_\a = e_{\ast}^{-V}  \ast {\boldnabla}_\a \ast e_{\ast}^V
\qquad , \qquad \nabla_{\adot} \equiv {\boldnabla}_{\adot} = \overline{D}_{\adot}
\eeq
with similar expressions for $(\overline{\nabla}_{\a}, \overline{\nabla}_{\adot})$.
These derivatives transform covariantly with respect to two types of
gauge transformations: quantum transformations
\bea
&& ~~~~~~~~~~~~~~~~~~~~~~~
e_{\ast}^V \rightarrow e_{\ast}^{i \overline{\L}} \ast e_{\ast}^V \ast
e_{\ast}^{-i \L}
\nonumber \\
&& \nabla_A \rightarrow e_{\ast}^{i\L} \ast \nabla_A \ast e_{\ast}^{-i\L}
\qquad , \qquad   {\boldnabla}_A \rightarrow  {\boldnabla}_A
\nonumber \\
&& \overline{\nabla}_A \rightarrow e_{\ast}^{i\overline{\L}} \ast \overline{\nabla}_A
\ast e_{\ast}^{-i\overline{\L}}
\qquad , \qquad   \overline{{\boldnabla}}_A \rightarrow  \overline{{\boldnabla}}_A
\label{quantum}
\eea
with background covariantly (anti)chiral parameters, ${\boldnabla}_\a \ast
\overline{\L} = {\boldnabla}_{\adot} \L = 0$, and background  transformations
\bea
&& ~~~~~~~~~~~~~~~~~~~~~~~~~~
e_{\ast}^V \rightarrow e_{\ast}^{i K} \ast e_{\ast}^V \ast
e_{\ast}^{-i K}
\nonumber \\
&& \nabla_A \rightarrow e_{\ast}^{iK} \ast \nabla_A \ast e_{\ast}^{-iK}
\qquad , \qquad   {\boldnabla}_A \rightarrow  e_{\ast}^{iK} \ast
{\boldnabla}_A \ast e_{\ast}^{-iK}
\nonumber \\
&& \overline{\nabla}_A \rightarrow e_{\ast}^{iK} \ast \overline{\nabla}_A
\ast e_{\ast}^{-iK}
\qquad , \qquad   \overline{{\boldnabla}}_A \rightarrow  e_{\ast}^{iK} \ast
\overline{{\boldnabla}}_A \ast e_{\ast}^{-iK}
\label{background}
\eea
with real parameter $K$.

Full covariantly (anti)chiral superfields
$\nabla_{\adot} \Phi = \nabla_{\a} \ast \overline{\Phi} = 0$
are expressed in terms of background
(anti)chiral superfields as $\Phi = \Phi_0$, $\overline{\Phi} = \overline{\Phi}_0
\ast e_{\ast}^V$, $\overline{\boldnabla}_{\adot} \ast \Phi_0 = 0$,
${\boldnabla}_{\a} \ast \overline{\Phi}_0 = 0$ and then linearly split
into a background and a quantum part. Under quantum transformations
the fields transform as $\Phi' = e_\ast^{i\L} \ast \Phi$,
$\overline{\Phi}' = \overline{\Phi} \ast e_\ast^{-i\overline{\L}}$, whereas under
background transformations they transform as $\Phi' = e_\ast^{iK} \ast \Phi$,
$\overline{\Phi}' = \overline{\Phi} \ast e_\ast^{-iK}$.

The classical action (\ref{action}) is invariant under the transformations (\ref{quantum},
\ref{background}). Background field quantization consists in performing
 gauge--fixing which explicitly breaks the (\ref{quantum}) gauge invariance
while preserving manifest invariance under (\ref{background}). 
The procedure follows closely the ordinary one \cite{superspace} by simply
replacing products with star products. It leads to a gauge--fixed action
$S_{tot} = S_{inv} + S_{GF} + S_{gh} $ where
$S_{gh}$ is given in terms of background covariantly (anti)chiral FP and NK ghost superfields and the quadratic part reads
\beq
S_{gh} =  \int d^4x d^4 \theta ~ \Big[ \overline{c}' c - c'\overline{c} + \overline{b} b \Big]
\label{ghosts}
\eeq 
From the rest of the action we read the $V$ propagator which in the Feynman gauge is 
\beq
\langle V_a(z) V_b(z') \rangle = g^2 \frac{\d_{ab}}{\Box_0} ~\d^{(4)}(\theta
-\theta')
\label{VVprop}
\eeq
and the pure gauge interaction terms useful for one--loop calculations 
\bea
&& -\frac{1}{2g^2} \int d^4x d^4 \theta~ {\rm Tr} V \Big[ -i [ \G^{\a \adot},
\pa_{\a \adot} V]_\ast - i \{ W^{\a} , D_{\a} V \}_\ast
- i \{ W^{\adot} , \overline{D}_{\adot}V \}_\ast
\nonumber \\
&&~~~~~~~~~ -\frac12 [ \G^{\a \adot} , [ \G_{\a \adot} , V]]_\ast -
\{ W^\a , [ \G_\a , V] \}_\ast
- \{ W^{\adot} , [ \G_{\adot} , V] \}_\ast \Big]
\label{VVvertex}
\eea

We now turn to the action for matter in a real representation
of the gauge group. In particular, ghosts fall in this category so the
following procedure can be applied to the action (\ref{ghosts}).

In the ordinary case, in terms of covariantly chiral and antichiral superfields
$\Phi$ and $\overline{\Phi}$
(related by complex conjugation)
\beq
S = \int d^4x d^4 \theta ~ \overline{\Phi} \Phi
\label{actionchiral}
\eeq

The corresponding equations of motion
\beq
{\cal O} \left( \begin{matrix}{ \Phi \cr \overline{\Phi} } \end{matrix} \right)
= 0
\quad \qquad
{\cal O} = \left( \begin{matrix} {0 & \overline{D}^2 \cr
                                 \nabla^2 & 0}
                  \end{matrix} \right)
\eeq
can be formally derived from the functional determinant
\beq
\Delta ~=~  \int {\cal D}\Psi
e^{ \overline{\Psi} {\cal O} \Psi }\sim  ({\rm det}{\cal O})^{-\frac{1}{2}}
\eeq
where $\Psi$ is the column vector $\left( \begin{matrix}{\Phi \cr \overline{\Phi}}
\end{matrix}\right)$.
If we perform the change of variables $\Psi = \sqrt{{\cal O}} \Psi'$,
whose jacobian is ${\rm det}\sqrt{{\cal O}} = \Delta^{-\frac{1}{2}}$, 
we can write
\beq
\Delta = \int {\cal D} \Psi'
~\Delta^{-1}~  e^{\overline{\Psi}' {\cal O}^2 \Psi'}
\label{Delta}
\eeq
or equivalently
\beq
\Delta^2 = \int {\cal D} \Psi
e^{\overline{\Psi} {\cal O}^2 \Psi}
\eeq
where
\beq
{\cal O}^2 = \left( \begin{matrix} {\overline{D}^2 \nabla^2 & 0 \cr
                                     0 & \nabla^2 \overline{D}^2}
                  \end{matrix} \right)
\label{calO}
\eeq
is a diagonal matrix. Therefore, defining the actions
\bea
&& S' = \frac12 \int d^4x d^4 \theta ~ \Phi \nabla^2 \Phi
\nonumber \\
&& \overline{S}' = \frac12 \int d^4x d^4 \theta ~ \overline{\Phi} \overline{D}^2 \overline{\Phi}
\label{actions}
\eea
it is easy to see that the following chain of identities holds
\cite{superspace}
\beq
\Delta^2 ~=~ \int {\cal D} \Phi {\cal D} \overline{\Phi}~ e^{S' + \overline{S}'}
~=~ \Big| \int {\cal D} \Phi e^{S'} \Big|^2
~=~ \Big( \int {\cal D} \Phi e^{S'} \Big)^2
\label{chain}
\eeq
where we have used $\overline{S}' = (S')^\dag$ and the fact that they
both contribute in the same way to $\Delta$ \cite{superspace}.
Therefore, when $\D$ is real,  we can identify the original
$\Delta$ with $\int {\cal D} \Phi e^{S'}$ and derive from here the Feynman
rules \cite{superspace}.

We now extend the previous derivation to the case of NAC euclidean superspace
where all the h.c. relations are relaxed and $\Phi$, $\overline{\Phi}$ are two
independent but {\em real} superfields.
 The matter action is still given by
(\ref{actionchiral}) and we can still define $\Delta$ as in (\ref{Delta}).
Therefore, we write
\beq
\Delta_\ast =
\int {\cal D} \Psi  ~
e^{\Psi^T \ast {\cal O}
\ast \Psi} \sim ({\rm det}({\bf O}))^{-\frac12}
\label{Delta2}
\eeq
 We can then proceed as before and square
the functional integral to obtain
\beq
\Delta_\ast^2 =  \int {\cal D} \Psi^T
e^{\Psi \ast {\bf O}^2 \ast \Psi}
\eeq
where ${\bf O}^2$ is given in (\ref{calO}) with the products promoted to
star products.
Now, if we introduce
\beq
\D_1 = \int {\cal D} \Phi   e^{S'}
\qquad , \qquad
\D_2 = \int {\cal D}\Phib e^{\overline{S}'}
\eeq
with $S'$, $\overline{S}'$ still given in (\ref{actions}) we can finally write
\beq
\D_{\ast}^2 = \D_1 \D_2
\label{r1}
\eeq
In contradistinction to the ordinary case, now $\overline{S}' \neq (S')^\dag$.
Moreover, the star product, when expanded,
could in principle generate different terms in the two actions. Therefore,
the chain of identities (\ref{chain}) is not immediately generalizable to the
NAC case and we cannot identify  $\D_{\ast} = \D_1$.

However, given the equality (\ref{r1}) the Feynman rules for $\D_{\ast}$ can be
still inferred from $\D_{1,2}$ order by order. In fact, we consider
$\D_{\ast}, \D_1, \D_2$ as functions of the coupling constant $g$ and perform
a perturbative expansion
\bea
\D_{\ast}[g] &&= \D_{\ast}[0] + g^2 \D_{\ast}'[0] + \cdots
\nonumber \\
\D_1[g] &&= \D_1[0] + g^2 \D_1'[0] + \cdots
\nonumber \\
\D_2[g] && = \D_2[0] + g^2 \D_2'[0] + \cdots
\eea
Normalizing the functionals as
$\D_{\ast}[0] =\D_1[0] =\D_2[0]$ and expanding the identity (\ref{r1})
in powers of $g$ we obtain
\beq
\D_{\ast}^2 = \big( 1 + g^2 \D_{\ast}'[0] + \cdots \big)^2 =
\big( 1 + g^2 \D_1'[0] + \cdots \big) \big( 1 + g^2 \D_2'[0] + \cdots \big)
\eeq
In particular, since we are interested in computing one--loop
contributions to the effective action at order $g^2$ we find
\beq \label{coef1/2}
2 \D_{\ast}'[0] = \D_1'[0] + \D_2'[0]
\eeq
Therefore at one loop $\D_{\ast}$ is given by the sum of the
contributions from $S'$ and $\overline{S}'$.

Following closely the ordinary case \cite{superspace} we derive the Feynman
rules from $S'$ and $\overline{S}'$ by first extracting the quadratic part of
the actions and then reading the vertices from the rest. 

Since for the chiral action the identities
involving covariant derivatives are formally the same except for 
the products which are now star products, the procedure to obtain the analytic
expressions
associated to the vertices is formally the same. We then refer the reader
to Ref. \cite{superspace} for details while reporting here only the final
rules:

\noindent
\begin{itemize}
\item
Propagator
\beq \label{prop}
\langle \Phi (z) \Phi(z') \rangle = -\frac{1}{\Box_0} \d^{(4)} (\theta -
\theta')
\eeq

\item
Chiral vertices: at one loop the prescription requires  associating with
one vertex
\beq
\frac12 \overline{D}^2 (\nabla^2 - D^2)
\label{v1}
\eeq
and with the other vertices
\beq
\frac12(\Box_+ - \Box_0)
\label{v2}
\eeq
where $\Box_+ ~=~ \Box_{cov} - i W^\a \ast \nabla_\a
- \frac{i}{2} (\nabla^\a \ast W_\a ), \qquad \Box_{cov}~=~ \frac{1}{2} \nabla^{\a \adot} \ast \nabla_{\a \adot}$.

\end{itemize}

The procedure can be easily extended to the case of massive chirals by
simply promoting the propagators (\ref{prop}) to massive propagators
$-1/(\Box_0 - m \overline{m})$. We also note that these rules are strictly one-loop rules.
At higher orders there are no difficulties and ordinary rules apply, as described in
\cite{superspace} with obvious modifications required by noncommutativity.

We can write down a formal effective interaction
lagrangian that corresponds to the one-loop rules above. In the  case of massive
matter (chirals with mass $m$ and antichirals with $\overline{m}$) in the adjoint
representation of the gauge group, it is given by (from now on we avoid indicating star products when no confusion arises)
\beq
S_0 + S_1 + S_2 \equiv \int d^4x d^4 \theta ~ {\rm Tr}
\left\{ \overline{\psi}(\Box_0 - m\overline{m})\psi + \frac{1}{2} \left[ \overline{\psi}
~\overline{D}^2 ( \Del^2 - D^2 )
\psi ~+~ \overline{\psi} (\Box_+ - \Box_0 ) \psi \right] \right\}
\label{effective}
\eeq
where $\psi, \overline{\psi}$ are {\em quantum unconstrained} superfields and the
first vertex has to appear once and only once in any one--loop diagram.\\
By writing everything explicitly in terms of connections and field strengths
and performing some integrations by parts it can be rewritten as
(we neglect terms with lower powers of $\overline{D}$
which would not contribute in one--loop calculations)
\bea
&& S_1 = \int d^4x d^4 \theta ~ {\rm Tr} \left\{ \left( \frac{i}{4} \G^\a
[ \overline{D}^2 \psi,
D_\a \overline{\psi} ]  - \frac{i}{4} \G^\a [ \overline{D}^2 D_\a \psi , \overline{\psi} ]
\right) +  \left( - \frac{1}{4} \overline{\psi} \{ \G^\a [ \G_\a, \overline{D}^2
\psi ]\} \right) \right\} \nonumber \\
&&~~~~\equiv S_1 + S'_1
\label{L} \\
&& S_2 = \int d^4x d^4 \theta ~ {\rm Tr} \left\{ \left( \frac{i}{4} \G^{\a\adot} [\psi,
\pa_{\a\adot} \overline{\psi} ]  - \frac{i}{4} \G^{\a \adot} [ \pa_{\a \adot} \psi , \overline{\psi} ] \right)
+ \left( \frac{i}{4} W^\a [\psi, D_\a \overline{\psi} ]
- \frac{i}{4} W^\a [ D_\a \psi , \overline{\psi} ] \right) \right.
\nonumber \\
&& \left.~~~~~~~~~~~~~~~~~~~~~~~~~~+~
\left( \frac14 [\G^\a , \psi][W_\a , \overline{\psi}]  
+ \frac14 [W^\a , \psi][\G_\a , \overline{\psi}] \right)
+ \left( \frac14 [\G^{\a \adot} ,\psi ][\G_{\a \adot}  ,\overline{\psi}] \right)
\right\}
\nonumber \\
&&~~~~\equiv S_2 + S'_2 + S_2'' + S_2'''
\label{R}
\eea

To extract the Feynman rules for the antichiral sector, 
we have to go carefully through the whole procedure since some of the identities 
used in the ordinary case do not hold anymore because of the noncommutative product.
We find convenient to express the action $\overline{S}'$ in terms of 
ordinary (not covariantly) antichiral superfields. Using cyclicity under trace and
$d^4 \theta$ integration we find
\beq
\overline{S}' = \frac12 \int d^4x d^4 \theta ~ {\rm Tr}(\overline{\Phi} e^{V} \overline{D}^2 
e^{-V} \overline{\Phi}) = \frac12 \int d^4x d^2 \overline{\theta} ~{\rm Tr} (\overline{\Phi} D^2 
\overline{\nabla}^2 \overline{\Phi}) 
\eeq
where $  \overline{\nabla}^2 =  e_{\ast}^{V} \ast \overline{D}^2 e_{\ast}^{-V}$. Using covariant derivatives 
in the antichiral representation we can formally follow 
the same procedure of the chiral sector by changing bar quantities into unbar ones, 
and viceversa. Therefore, the Feynman rules are:

\begin{itemize}
\item Propagator

\beq
\langle \overline{\Phi} (z) \overline{\Phi}(z') \rangle =
-\frac{1}{\Box_0} \d^{(4)} (\theta - \theta')
\eeq

\item one vertex: $\frac12 D^2 (\overline{\nabla}^2 - \overline{D}^2)$

\item other vertices: $\frac12(\overline{\Box}_+ - \Box_0)$

with $\overline{\Box}_+ = \overline{\Box}_{cov} - i\overline{W}^{\adot} \ast \overline{\nabla}_{\adot} - \frac{i}{2}
( \overline{\nabla}^{\adot} \ast \overline{W}_{\adot})$ (here $\overline{W}_{\adot}$ is the antichiral field strength and $\overline{\Box}_{cov}~=~ \frac{1}{2} \overline{\nabla}^{\a \adot} \ast \overline{\nabla}_{\a \adot}$).

\end{itemize}

Again, it is convenient to introduce an effective action in terms of quantum
unconstrained superfields $\xi$ and $\overline{\xi}$
\beq
\overline{S}_0 + \overline{S}_1 + \overline{S}_2 \equiv \int d^4x d^4 \theta ~ {\rm Tr}
\left\{ \overline{\xi}(\Box_0 - m\overline{m})\xi + \frac{1}{2} \left[ \overline{\xi}
D^2 (\overline{\nabla}^2 - \overline{D}^2)
\xi ~+~ \overline{\xi} (\overline{\Box}_+ - \Box_0 ) \xi \right] \right\}
\label{effective-b}
\eeq
In terms of connections and field strengths it can be rewritten as
\bea
&& \overline{S}_1 = \int d^4x d^4 \theta ~ {\rm Tr} \left\{ \left( \frac{i}{4} \overline{\G}^{\adot}
[ \xi,
\overline{D}_{\adot} D^2 \overline{\xi} ]  - \frac{i}{4} \overline{\G}^{\adot} [ \overline{D}_{\adot} \xi , D^2 \overline{\xi} ] \right) +  \left( - \frac{1}{4} \overline{\xi} \{ \overline{\G}^{\adot} [ \overline{\G}_{\adot}, D^2
\xi ]\} \right) \right\} \nonumber \\
&&~~~~\equiv \overline{S}_1 + \overline{S}'_1 \label{Lbar} \\
&& \overline{S}_2 = \int d^4x d^4 \theta ~ {\rm Tr} \left\{ \left( \frac{i}{4} \overline{\G}^{\a \adot} [\xi,\pa_{\a \adot} \overline{\xi} ]  - \frac{i}{4} \overline{\G}^{\a \adot} [ \pa_{\a \adot} \xi , \overline{\xi} ] \right) + \left( \frac{i}{4} \overline{W}^{\adot} [\xi, \overline{D}_{\adot} \overline{\xi} ] - \frac{i}{4} \overline{W}^{\adot} [ \overline{D}_{\adot} \xi , \overline{\xi} ] \right) \right.
\nonumber \\
&& \left.~~~~~~~~~~~~~~~~~~~~~~~~~~+~
\left( \frac14 [\overline{\G}^{\adot} , \xi][\overline{W}_{\adot} , \overline{\xi}]  
+ \frac14 [\overline{W}^{\adot} , \xi][\overline{\G}_{\adot} , \overline{\xi}] \right)
+ \left( \frac14 [\overline{\G}^{\a\adot} ,\xi ][\overline{\G}_{\a \adot}  ,\overline{\xi}] \right)
\right\}
\nonumber \\
&&~~~~\equiv \overline{S}_2 + \overline{S}'_2 + \overline{S}_2'' + \overline{S}_2'''
\label{Rbar}
\eea


\section{Supergraph techniques in momentum frame}

As in the ordinary supersymmetric theories, it is convenient to develop a
procedure which allows one to study NAC field theories perturbatively without
going to components, in particular without expanding necessarily the star 
product.
As already indicated in the Introduction, such a procedure is
unavoidable in dealing with NAC super Yang--Mills theories when the relation
between nonaticommutativity and gauge invariance is under study.

In this section we describe the main lines of our approach. Details will be
reported in a future publication \cite{GPR2}.

Given a generic superfunction $\Phi$ living on the NAC superspace we move to
{\em momentum superspace} by
Fourier transforming both  the bosonic and fermionic coordinates
according to the prescription
\beq
\widetilde{\Phi}(p, \pi, \overline{\pi}) = \int d^4x d^2 \theta d^2 \overline{\theta}~
e^{ipx + i\pi \theta + i\overline{\pi} \overline{\theta}}~ \Phi(x, \theta, \overline{\theta})
\label{completeFT}
\eeq
In momentum superspace the star product is
traded for an exponential factor dependent on the spinorial momentum
variables
\beq
\Phi (x,\theta, \overline{\theta}) \ast \Psi(x,\theta, \overline{\theta})
\longrightarrow e^{\pi \wedge \pi'} ~\widetilde{\Phi}(p,\pi,\overline{\pi})
\widetilde{\Psi}(p',\pi',\overline{\pi}')
\eeq
where we have defined $\pi \wedge \pi' \equiv \pi_\a {\cal F}^{\a \b} \pi'_\b$.

We then develop perturbative techniques in momentum superspace. In the ordinary
anticommutative case this amounts to translating Feynman rules for propagators
and vertices to momentum language. Spinorial $D$ derivatives
become $\tilde{D}$ derivatives according to the relations 
\bea
&& D_{\alpha} = \partial_{\alpha} + i \overline{\theta}^{\dot{\alpha}}
\partial_{\alpha \dot{\alpha}}
\quad \rightarrow  \qquad\widetilde{D}_{\alpha} = - i \pi_{\alpha} -i
\overline{\eth}^{\dot{\alpha}} p_{\alpha \dot{\alpha}} \nonumber \\
&& \overline{D}_{\dot{\alpha}} = \overline{\partial}_{\dot{\alpha}} \qquad
\quad\qquad \rightarrow
\qquad \widetilde{\overline{D}}_{\dot{\alpha}} = - i \overline{\pi}_{\dot{\alpha}}
\label{Dtilde}
\eea
where we have indicated $\overline{\eth}_{\dot{\alpha}} = \frac{\pa}{\pa \overline{\pi}^{\adot}}$.
The ordinary $D$--algebra which allows  reducing a supergraph to an ordinary
momentum diagram gets translated into a $\tilde{D}$--algebra in a
straightforward way. In particular, while in configuration superspace the
general rule to get a notrivial contribution from a given supergraph is to
perform $D$--algebra until we are left with a factor $D^2\overline{D}^2$ for
each loop, in momentum superspace it gets translated into the requirement to
perform $\tilde{D}$--algebra until one ends up with a factor
$\pi^2 \overline{\pi}^2$ for each loop, where $(\pi, \overline{\pi})$ are the loop
spinorial momenta.

In the NAC case we again translate the Feynman rules to momentum superspace.
However, relevant changes occur due to the appearance of exponential
factors at the vertices. Thus, given a local cubic vertex of the form
$\int A \ast B \ast C$ the corresponding expression in momentum superspace
becomes (we indicate $\Pi \equiv (p,\pi,\overline{\pi})$)
\beq
e^{\p_1 \wedge \pi_2}~\tilde{A}(\Pi_1)\tilde{B}(\Pi_2)\tilde{C}(\Pi_3)
~\d^{(8)}(\Pi_1 + \Pi_2 + \Pi_3)
\eeq
More generally a local $n$--point vertex gives
\beq
\prod_{i <j }^n ~e^{\p_i \wedge \pi_j} ~\tilde{A}_1(\Pi_1) \cdots
\tilde{A}_n(\Pi_n)~\d^{(8)}(\sum_i\Pi_i)
\eeq
When contracting quantum lines coming out from the vertices to build Feynman
diagrams, different ways of performing contractions lead to different
configurations of exponential factors. Due to spinorial momentum conservation
at each vertex the diagrams can be classified into {\em planar} diagrams
characterized by exponentials depending only on the external momenta and
{\em nonplanar} ones which have a  nontrivial exponential dependence on the
loop momenta. This pattern resembles closely what happens in the case
of bosonic noncommutative theories \cite{filk, minwalla} except for the
exponential factors which in that case are actual phase factors. However,
this does not prevent us from using the same prescriptions 
\cite{filk, minwalla}
to determine the overall exponential factor associated with a given diagram.

Once the exponential factor and the structure of the $\tilde{D}$ derivatives
associated to a given diagram have been established we proceed by performing
$\tilde{D}$--algebra. This amounts to using suitable identities to reduce the
number of spinorial derivatives, expanding the exponential factors as
\beq
 e^{\pi_1 \wedge \pi_2} = 1 + \pi_1^\a {\cal F}_{\a\b} \pi_2^\b -
\frac12 \pi_1^2 {\cal F}^2 \pi_2^2
\label{exp}
\eeq
and selecting those configurations of spinorial momenta which have a factor
$\pi^2 \overline{\pi}^2$ for each loop. We note that while $\overline{\pi}^2$ factors
only come from $\tilde{\bar D}$ derivatives associated to the vertices as in
the ordinary case, $\pi^2$ factors can also come from the expansion (\ref{exp}),
giving extra nonvanishing contributions to a given diagram proportional to
the nonanticommutation parameter ${\cal F}$. This is the way
nonanticommutativity enters the calculations in our approach.

Finally, once $\tilde{D}$--algebra has been performed, we are left with
ordinary momentum loop integrals. We evaluate them in dimensional
regularization ($n = 4 - 2\e$) and in the G--scheme \cite{gscheme} in order
to avoid dealing with irrelevant constants coming from the expansion of gamma
functions.

Before closing this section we give the Feynman rules in momentum superspace
needed for calculations at one--loop in the background field approach.

The propagators for the gauge superfield and massive matter
are (see eqs. (\ref{VVprop}, \ref{effective}))
\bea
&&\langle \tilde{V}^a(\Pi) \tilde{V}^b(\Pi') \rangle =
-g^2 \frac{\d^{ab}}{p^2} ~\d^{(8)}(\Pi + \Pi')
\nonumber \\
&& \langle \tilde{\psi}(\Pi) \tilde{\overline{\psi}}(\Pi') \rangle =
\frac{1}{p^2 + m \overline{m}}  ~\d^{(8)}(\Pi + \Pi')
\label{FTprop}
\eea
where we take the momenta always entering the vertex.

The vertices quadratic in the quantum $V$ superfield are given in
(\ref{VVvertex}). The only term which effectively contributes will be the one
which contains a $\overline{D}$.
Performing Fourier transform as described above the corresponding vertex gets the structure (we indicate $\Pi_i \equiv (p_i, \pi_i,
\overline{\pi}_i))$)
\bea
&& \frac{1}{2 g^2} \int d^8 \Pi_1  d^8 \Pi_2 d^8 \Pi_3
~\d^{(8)} (\Pi_1 + \Pi_2 +\Pi_3) ~\widetilde{W}^{\ad \, a} (\Pi_1)
\overline{\pi}_{2 \, \ad}
\tilde{V}^b(\Pi_2) \tilde{V}^c(\Pi_3)
\nonumber \\
&&~~~~~~~~~~~~~\times \Big[ {\rm Tr}(T^a [ T^b,T^c])
\cosh{(\pi_1 \wedge \pi_2)}
+ {\rm Tr} (T^a \{ T^b , T^c\}) \sinh{(\pi_1 \wedge \pi_2)} \Big]
\label{vertex2}
\eea
where we use the convention that all the momenta are incoming.

We now consider the gauge--matter vertices. Their structures can be read from the effective actions (\ref{effective}) and (\ref{effective-b}). Performing FT of the terms which eventually contribute in a nontrivial way, for the cubic vertices in the chiral sector we find
\bea
&& \tilde{S}_1 =  \frac{i}{4} \int d^8 \Pi_1 d^8 \Pi_2 d^8 \Pi_3 ~
\d^{(8)} ( \Pi_1 + \Pi_2 + \Pi_3 )~ \overline{\pi}_2^2
~\widetilde{\G}^{\a}_a(\Pi_1)
\nonumber \\
~&\times&  \big[ ( -i\pi_{2\a} - i
\overline{\eth}_2^{\adot} p_{2 \,\a\adot}) \psi_b(\Pi_2) \overline{\psi}_c (\Pi_3)
- \psi_b( \Pi_2) (-i\pi_{3\a} - i
\overline{\eth}_3^{\adot} p_{3 \, \a\adot}) \overline{\psi}_c (\Pi_3) \big]
\nonumber \\
~&\times& \left\{ {\rm Tr} (T^a [ T^b , T^c ]) \cosh{(\pi_1 \wedge \pi_2)}
+ {\rm Tr} (T^a \{ T^b , T^c \}) \sinh{(\pi_1 \wedge \pi_2)} \right\}
\label{tildeS1}
\eea
\bea
\tilde{S}_2 &=&  \frac{1}{4} \int d^8 \Pi_1 d^8 \Pi_2 d^8
\Pi_3 ~ \d^{(8)} ( \Pi_1 + \Pi_2 + \Pi_3 )
~\widetilde{\G}^{\mu}_a(\Pi_1)
~  [-p_{2 \,\mu} + p_{3 \,\mu}] \psi_b(\Pi_2) \overline{\psi}_c (\Pi_3)
\nonumber \\
&\times& \left\{ {\rm Tr} (T^a [ T^b , T^c ]) \cosh{(\pi_1 \wedge \pi_2)}
+ {\rm Tr} (T^a \{ T^b , T^c \}) \sinh{(\pi_1 \wedge \pi_2)} \right\}
\label{tildeS2}
\eea
\bea
&& \tilde{S}'_2 = -\frac{i}{4} \int d^8 \Pi_1 d^8 \Pi_2 d^8
\Pi_3 ~\d^{(8)} ( \Pi_1 + \Pi_2 + \Pi_3 )
~\widetilde{W}^{\a}_a(\Pi_1)
\nonumber \\
~&\times&  \big[ ( -i\pi_{2\a} - i
\overline{\eth}_2^{\adot} p_{2 \, \a\adot}) \psi_b(\Pi_2) \overline{\psi}_c (\Pi_3)
-\psi_b(\Pi_2) (-i\pi_{3\a} - i
\overline{\eth}_3^{\adot} p_{3 \, \a\adot}) \overline{\psi}_c (\Pi_3) \big]
\nonumber \\
~&\times& \left\{ {\rm Tr} (T^a [ T^b , T^c ]) \cosh{(\pi_1 \wedge \pi_2)}
+ {\rm Tr} (T^a \{ T^b , T^c \}) \sinh{(\pi_1 \wedge \pi_2)} \right\}
\label{tildeS2'}
\eea
whereas the quartic vertices are given by
\bea
&& \tilde{S}'_1 =  \frac{1}{4}\int d^8 \Pi_1 d^8 \Pi_2 d^8 \Pi_3 d^8 \Pi_4~
\d^{(8)} ( \Pi_1 + \Pi_2 + \Pi_3 + \Pi_4)~ \overline{\pi}_4^2 \nonumber \\
&&~~~~~~~~~~\times ~\overline{\psi}_a(\Pi_1)~
\widetilde{\G}^{\a}_b(\Pi_2)~\widetilde{\G}_{c~\a}(\Pi_3)~
\psi_d (\Pi_4)
\nonumber \\
&& ~~~~~~~~~~ \times ~ {\cal P}^{abcd} ( \pi_1 \pi_2 \pi_3 \pi_4)
\label{tildeS1'}
\\
&& \tilde{S}''_2 =  -\frac{1}{4}\int d^8 \Pi_1 d^8 \Pi_2 d^8 \Pi_3 d^8 \Pi_4~
\d^{(8)} ( \Pi_1 + \Pi_2 + \Pi_3 + \Pi_4) \nonumber \\
&&~~~~~~~~~~\times ~\overline{\psi}_a(\Pi_1)
\left( \widetilde{W}^{\a}_b(\Pi_2)~\widetilde{\G}_{c~\a}(\Pi_3) +
\widetilde{\G}^{\a}_b(\Pi_2)~\widetilde{W}_{c~\a}(\Pi_3) \right)~
\psi_d (\Pi_4) \nonumber \\
&& ~~~~~~~~~~ \times ~ {\cal P}^{abcd} ( \pi_1 \pi_2 \pi_3 \pi_4)
\label{tildeS2''}
\\
&& \tilde{S}'''_2 =  -\frac{1}{4}\int d^8 \Pi_1 d^8 \Pi_2 d^8 \Pi_3 d^8 \Pi_4~
\d^{(8)} ( \Pi_1 + \Pi_2 + \Pi_3 + \Pi_4) \nonumber \\
&&~~~~~~~~~~\times ~\overline{\psi}_a(\Pi_1)~
\widetilde{\G}^{\a\adot}_b(\Pi_2)~\widetilde{\G}_{c~\a\adot}(\Pi_3) ~
\psi_d (\Pi_4)
\nonumber \\
&& ~~~~~~~~~~ \times ~{\cal P}^{abcd} ( \pi_1 \pi_2 \pi_3 \pi_4)
\label{tildeS2'''}
\eea
with 
\bea
{\cal P}^{abcd} ( \pi_1 \pi_2 \pi_3 \pi_4) &=& \big\{ ~{\rm Tr}([T^a,T^b][T^c,T^d])~\cosh{(\pi_1 \wedge \pi_2)}~
\cosh{(\pi_3 \wedge \pi_4)} \nonumber \\
&& ~~~+ {\rm Tr}(\{T^a,T^b\}[T^c,T^d])~
\sinh{(\pi_1 \wedge \pi_2)}~\cosh{(\pi_3 \wedge \pi_4)} \nonumber \\
&& ~~~+ {\rm Tr}([T^a,T^b]\{T^c,T^d\})~
\cosh{(\pi_1 \wedge \pi_2)}~
\sinh{(\pi_3 \wedge \pi_4)}\nonumber \\
&& ~~~+ {\rm Tr}(\{T^a,T^b\}\{T^c,T^d\})~
\sinh{(\pi_1 \wedge \pi_2)}~
\sinh{(\pi_3 \wedge \pi_4)} \big\} \nonumber
\eea
For the antichiral vertices we have analogous expressions with obvious 
changes.


\section{One--loop diagrams}

In this section, using the techniques described above, we compute the one--loop divergent 
contributions to the gauge effective action. In the ordinary case, in background field method
only the two--point function with chiral loop is divergent \cite{GSZ, superspace}. In the NAC
case, instead, we find divergent contributions up to the 4--point function for the gauge field
due to the nontrivial modifications to the $D$--algebra induced by the star product.

We list our results without details. These will be reported elsewhere \cite{GPR2}.
All the divergences are expressed in terms of a tadpole integral ${\cal T}$ and
a self--energy ${\cal S}$ which in dimensional regularization ($n = 4 - 2\e$) are
\bea
&& {\cal T} ~~\equiv \int d^4 q ~\frac{1}{q^2 + m \overline{m}}  
= - \frac{m \overline{m}}{(4\pi)^2}~\frac{1}{\epsilon} + {\cal O}(1)
\label{tadpole}\\
&& {\cal S} \equiv
\int d^4 q ~\frac{1}{((q-p)^2 + m \overline{m})(q^2 + m \overline{m})}
= \frac{1}{(4\pi)^2}~\frac{1}{\epsilon} + {\cal O}(1)
\label{selfenergy}
\eea
Other one--loop divergent integrals are obtained in terms of ${\cal T}$
and ${\cal S}$ through the following identities
\beq
\int d^4 q ~\frac{q_{\a \adot}}{((q-p)^2 + m \overline{m})(q^2 + m \overline{m})}
= \frac12 p_{\a\adot} ~{\cal S}
\label{S1}
\eeq
\bea
&& \int d^4 q ~\frac{q_{\a \adot}q_{\b \dot{\b}}}{((q-p)^2 + m \overline{m})
(q^2 + m \overline{m})} =
\nonumber \\
&&~~~~~
\frac13 C_{\a\b} C_{\adot \dot{\b}} \left[ {\cal T} - \frac12 (p^2 + 4 m
\overline{m}) {\cal S}
\right]
~+~ \frac13 \frac{p_{\a\adot} p_{\b \dot{\b}}}{p^2} \left[
{\cal T} + (p^2 + m \overline{m}) {\cal S} \right]
\label{S2}
\eea
\beq
\int d^4 q ~\frac{q^2}{((q-p)^2 + m \overline{m})
(q^2 + m \overline{m})} = {\cal T} - m \overline{m} {\cal S}
\label{S3}
\eeq
\beq \label{S4}
\int d^4 q \frac{q_{\a\adot} q_{\b \bdot}}{(q^2 + m\overline{m})( (q+p)^2 + m\overline{m})
((q+r)^2 + m\overline{m})} \sim
\frac{1}{2} C_{\a\b} C_{\adot \bdot} ~ {\cal S} \\
\eeq
\bea \label{S5}
&& \int d^4 q \frac{q_{\a\adot} q_{\b \bdot}q_{\g\gdot} q_{\rho \dot{\rho}} }{(q^2 + m\overline{m})( (q+p)^2 + m\overline{m})
((q+r)^2 + m\overline{m})((q+s)^2 + m\overline{m})} \sim \nonumber \\ 
&& \qquad \qquad \frac{1}{6} {\cal S} \big( C_{\a\b} C_{\adot \bdot} C_{\g\rho} C_{\gdot \dot{\rho}} + C_{\a\g} C_{\adot \gdot} C_{\b\rho} C_{\bdot \dot{\rho}} + C_{\a\rho} C_{\adot \dot{\rho}} C_{\b\g} C_{\bdot \gdot} \big) \nonumber \\
\eea

In the massless case ($m = \overline{m} =0$) the tadpole ${\cal T}$ vanishes due to
a complete cancellation between the UV and the IR divergence. Consequently,
the results
for the self--energy type integrals can be obtained from (\ref{selfenergy} - \ref{S5}) by setting ${\cal T} \sim 0$ and $m = \overline{m} =0$.


\subsection{Two--point function}

Divergent contributions to the two--point function are listed in Fig. 1.

\begin{minipage}{\textwidth}
\begin{center}
\includegraphics[width=0.80\textwidth]{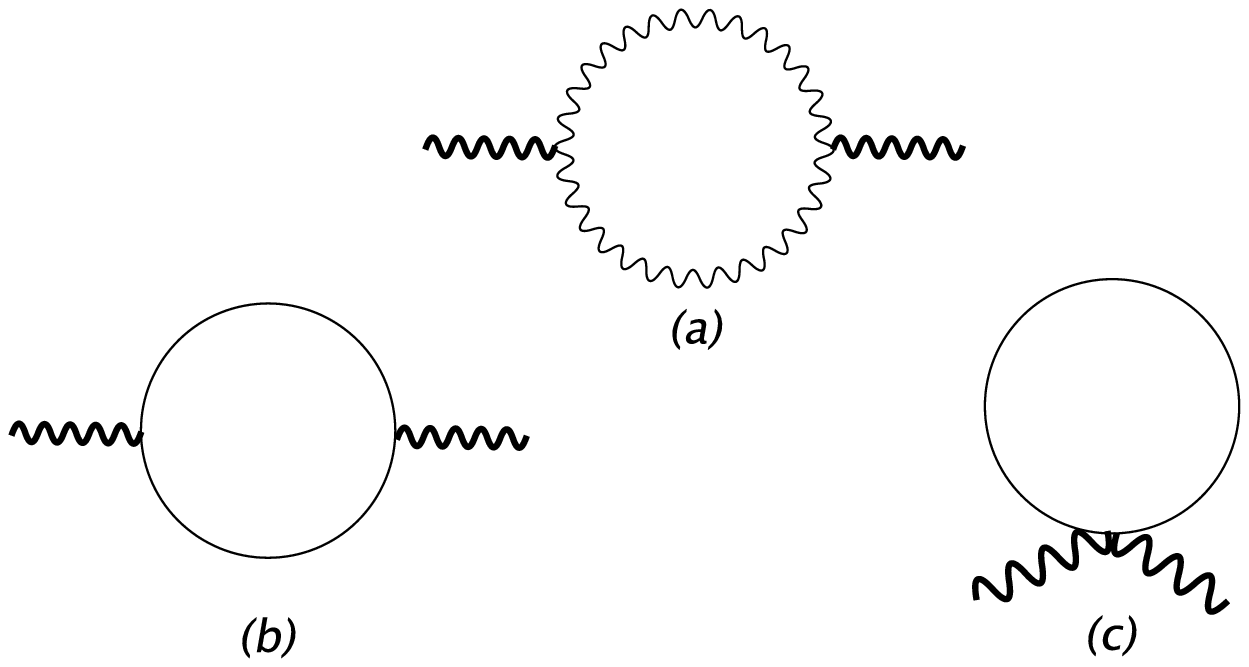}
\end{center}
\begin{center}
{\small{Figure 1: Gauge one--loop two--point functions. }}
\end{center}
\end{minipage}

Diagram (1a) with a vector loop can be computed by using Feynman rules (\ref{FTprop}) and (\ref{vertex2}).
The divergent contribution turns out to be proportional to ${\cal F}^2$
\beq
-2{\cal F}^2 \int d^4 x d^4 \theta ~ {\rm Tr}(\pa^2 W^\ad)
{\rm Tr}(W_{\ad})
=-2{\cal F}^2
\int d^4 x d^4 \theta ~ \pa^2 {\rm Tr}(e_\ast^{-V} \ast
\overline{W}^\ad
\ast e_\ast^V) {\rm Tr}(e_\ast^{-V} \ast \overline{W}_{\ad} \ast e_\ast^V)
\label{Vresult}
\eeq
where we have expressed the covariantly antichiral field strength
$W_\ad$ in terms of the ordinary antichiral one.\\
Given the particular group structure one can prove that this contribution
is equal to
\beq
-2{\cal F}^2 ~{\cal S}~\int d^4 x d^4 \theta ~ (\pa^2 {\rm Tr}\overline{W}^\ad)
\Big[ {\rm Tr}\overline{W}_{\ad} + 2 {\cal F}^{\a\b}{\rm Tr}( \pa_\a e_\ast^V \pa_\b
(e_\ast^{-V} \ast \overline{W}_\ad)) \Big]
\eeq
and actually vanishes once integrated in $d^2 \theta$.\\
\\
We now consider matter loops (1b,1c) following the Feynman rules (\ref{FTprop}) and (\ref{tildeS1})--(\ref{tildeS2'''}) for the chiral superfields and the analogous ones for the antichirals. This also covers contributions from ghosts up to an overall sign. \\
We focus on the chiral and the antichiral sectors separately. \\
\\
$\bullet$ Chiral sector:\\
\noindent
Order ${\cal F}^0$: This is the contribution which is already present in the
the ordinary case. It is obtained by taking the $\pi^2$ factor inside the loop entirely from
the covariant derivatives. Performing the explicit calculation we find
\beq
{\cal D}^{(2)}= \frac{1}{2}{\cal S} \int d^8 z \Big[~{\cal N}~{\rm Tr}\left(\Gamma^{\a} ~W_{\a}\right)   - {\rm Tr}\left( \Gamma^{\a}  \right){\rm Tr}\left(  W_{\a}\right)\Big]
\eeq

\noindent
Order ${\cal F}$:
Contributions proportional to a single power of ${\cal F}$ come from
diagrams which have already a single power $\pi$ from the covariant
derivatives, whereas a second factor $\pi$ is produced by linearly
expanding the hyperbolic functions. However, it is easy to realize that
 this expansion is always proportional to a trivial vanishing colour factor.

\noindent
Order ${\cal F}^2$:
These contributions are associated to nonplanar diagrams where
no $\pi$ factors come from covariant derivatives and the hyperbolic functions
are expanded up to  second order. After a bit of calculations we obtain
\bea \label{result2}
&&\frac{{\cal F}^2}{4}~\int d^8z ~\Big\{ - {\rm Tr}\left((4{\cal T} -
4m \overline{m} {\cal S} + \Box ~{\cal S}) \pa^2 \frac{D^2}{\Box}\G^\a \right)
{\rm Tr}(\overline{D}^2\G_\a) \nonumber \\
&& \qquad \qquad \qquad + \frac{2}{3} i {\rm Tr}\left( ( 2{\cal T} - 4 m
\overline{m}{\cal S} + \Box~{\cal S} ) \pa^2 D_{\beta} \frac{\pa^{\beta \dot{\a}}}
{\Box}\G^{\a} \right){\rm Tr}(\overline{D}_{\dot{\a}}\G_{\a})\nonumber \\
&& \qquad \qquad \qquad + \frac{i}{3} {\rm Tr}\left( ( 4{\cal T} + 4 m \overline{m}
{\cal S} - \Box~{\cal S}) \frac{D^{\a}}{\Box}~\pa^2 \G_\a\right){\rm Tr}
(\pa^{\beta\dot{\beta}}\overline{D}_{\dot{\beta}}\G_{\beta})\nonumber \\
&& \qquad \qquad \qquad - 4~ {\cal T}~{\rm Tr}( \pa^2 \G^{\a} )~{\rm Tr}
( \G_{\a})\Big\}
\eea
It is possible to prove that this expression actually vanishes due to the particular structure of the star product hidden in $\G_\a$ and the fact that
in (\ref{result2}) only the $U(1)$ part of the connection appears.\\
\\
$\bullet$ Antichiral sector:\\
\noindent
Order ${\cal F}^0$: Also in this case this is the contribution which is already present in the ordinary case. 
\beq
\overline{{\cal D}}^{(2)}= \frac{1}{2}{\cal S} \int d^8 z \Big[~{\cal N}~{\rm Tr}\left(\overline{\Gamma}^{\adot} ~\overline{W}_{\adot}\right)   - {\rm Tr}\left( \overline{\Gamma}^{\adot}  \right){\rm Tr}\left(  \overline{W}_{\adot}\right)\Big]
\eeq
\noindent
Order ${\cal F}$: As in the chiral sector, this contribution is proportional to a trivial vanishing colour factor, then there is no contribution at this order.

\noindent
Order ${\cal F}^2$: Divergent contributions proportional to ${\cal F}^2$ have the form  
\beq
{\cal F}^2 \int d^8 z \overline{\theta}^2 ~\Box~ \pa^2 \overline{\Gamma}^{\adot}~ \overline{W}_{\adot} = {\cal F}^2 \int d^8 z \overline{\theta}^2 ~\Box ~D^2 \overline{\Gamma}^{\adot} ~\overline{W}_{\adot} 
\eeq    

and trivially vanish when integrated in $d^2 \theta$.


\subsection{Three--point function}

Divergent contributions to the three--point function are listed in Fig. 2.

\begin{minipage}{\textwidth}
\begin{center}
\includegraphics[width=0.60\textwidth]{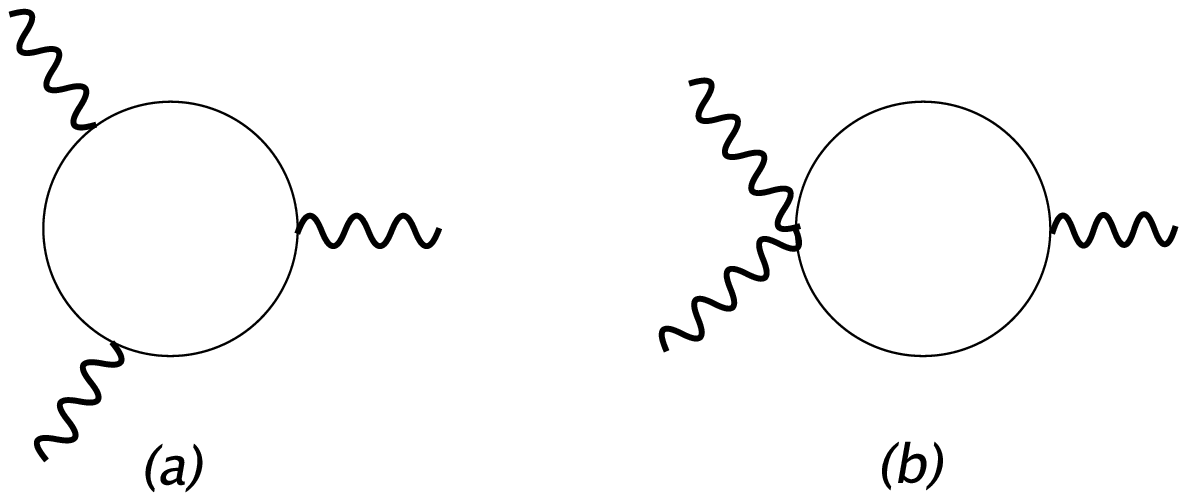}
\end{center}
\begin{center}
{\small{Figure 2: Gauge one--loop three--point functions. }}
\end{center}
\end{minipage}

Three-point diagrams with vector loop are finite, then we focus on matter loops. Both in the chiral and in the antichiral sectors, divergent terms are of order ${\cal F}$. Indeed, even if in principle we can have divergences also at order ${\cal F}^2$, the explicit calculation of the phase structures shows that these contributions cancel.\\
$\bullet$ Chiral sector:\\
We obtain
\bea
{\cal D}^{(3)}&=& i {\cal S} \int d^4 x d^2 \theta ~ {\cal F}^{\b\g} [ {\rm Tr}(\pa_\b W^\a) \ast 
{\rm Tr}(\G_\g  \ast W_\a) - {\rm Tr}(\pa_\b W^\a) \ast 
{\rm Tr}(W_\a \ast \G_\g )
\nonumber \\
&& \qquad \qquad \qquad \qquad\qquad \qquad \qquad - {\rm Tr}(\pa_\b \G_\g) \ast {\rm Tr}(W^\a  \ast W_\a) ]_{\overline{\theta}=0}
\label{3ptch1}
\eea
We have already performed the $\overline{\theta}$ integration since it allows for
some cancellation among various terms.
One can prove that the star products in (\ref{3ptch1}) are actually ordinary
products and the result can be rewritten as
\beq
{\cal D}^{(3)}= i {\cal S} \int d^4 x d^2 \theta ~ {\cal F}^{\b\g} [ 2{\rm Tr}(D_\b W^\a)
{\rm Tr}(\G_\g   W_\a)- {\rm Tr}(D_\b \G_\g) 
{\rm Tr}(W^\a W_\a) ]_{\overline{\theta}=0}
\label{3ptch2}
\eeq
$\bullet$ Antichiral sector:\\
We obtain 
\bea
\overline{{\cal D}}^{(3)} &=& \frac{2}{3} {\cal S} {\cal F}^{\rho \g} \int d^8 z~\overline{\theta}^{\bdot}\Big[ {\rm Tr}\left( \pa_{\rho} \overline{\Gamma}^{\adot} \right){\rm Tr}\left( \overline{W}_{\adot}\overline{\Gamma}_{\g \bdot}\right) + {\rm Tr}\left( \pa_{\rho} \overline{W}^{\adot}\right){\rm Tr}\left( \overline{\Gamma}_{\adot} \overline{\Gamma}_{\g \bdot}\right)\nonumber \\
&& \qquad \qquad \qquad \qquad\qquad \qquad \qquad + {\rm Tr}\left( \pa_{\rho} \overline{\Gamma}_{\g \bdot}\right){\rm Tr}\left( \overline{W}^{\adot} \overline{\Gamma}_{\adot}\right) \Big]
\eea
In this case we perform both the $\theta$ and $\overline{\theta}$ integrations, in order to allow for some cancellations 
\beq
\overline{{\cal D}}^{(3)} = i {\cal S} {\cal F}^{\rho \g} \int d^4 x \Big[~2~\pa_{\rho \dot{\rho}}{\rm Tr}\left( \overline{W}^{\adot} \right) {\rm Tr}\left(\overline{W}_{\adot} \overline{\Gamma}_{\g}^{~\dot{\rho}} \right) + \pa_{\rho \dot{\rho}}{\rm Tr}\left( \overline{\Gamma}_{\g}^{~\dot{\rho}} \right) {\rm Tr}\left(\overline{W}^{\adot}\overline{W}_{\adot}\right)\Big]_{\theta=\overline{\theta}=0}
\eeq


\subsection{Four--point function}

Divergent contributions to the four--point function are listed in Fig. 3.

\begin{minipage}{\textwidth}
\begin{center}
\includegraphics[width=0.60\textwidth]{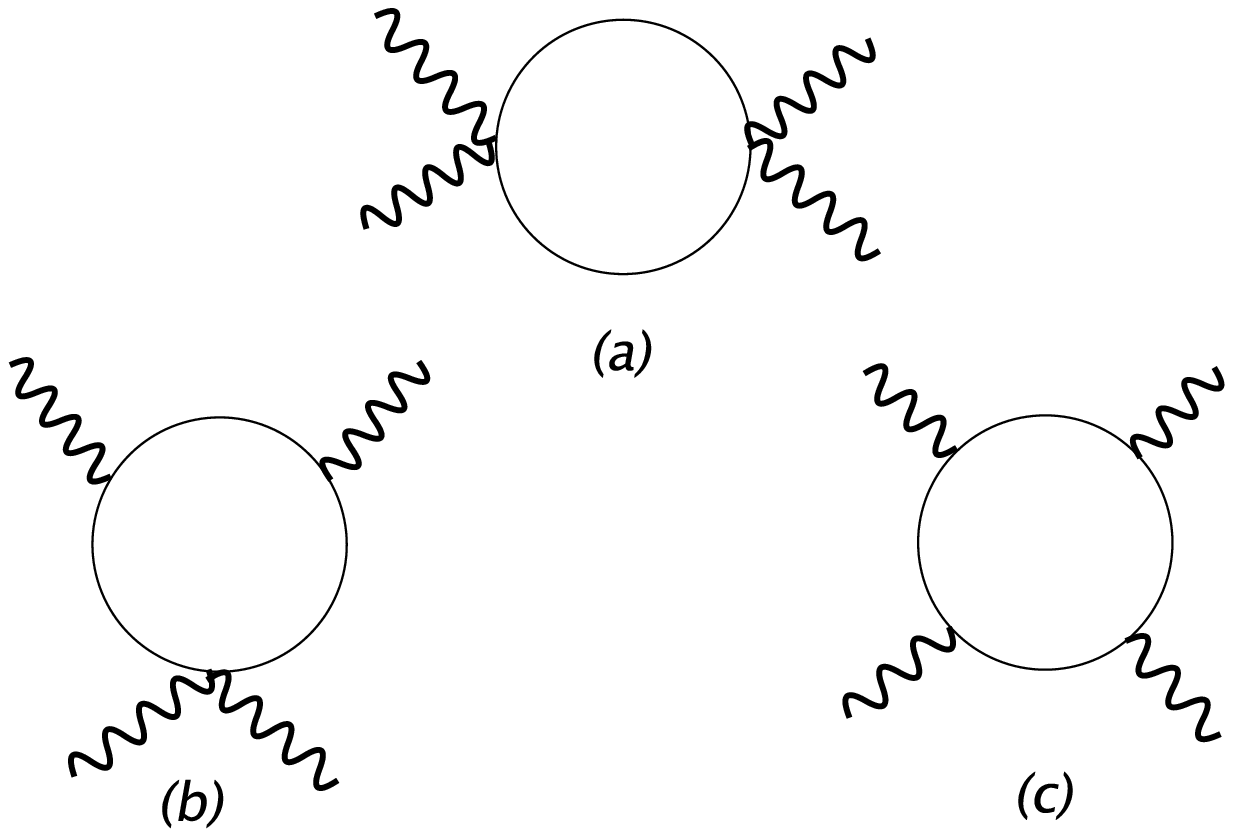}
\end{center}
\begin{center}
{\small{Figure 3: Gauge one--loop four--point functions. }}
\end{center}
\end{minipage}

Again, diagrams with vector loops give finite contributions.
Considering matter loops we find divergences only at order ${\cal F}^2$.\\
$\bullet$ Chiral sector:\\
After $\overline{\theta}$ integration, we obtain 
\bea
{\cal D}^{(4)}&=& {\cal S} {\cal F}^2 \int d^4 x d^2 \theta ~\Big[ \frac{1}{2} \pa^2{\rm Tr} \left( \Gamma^{\a}\ast\Gamma_{\a} \right)  {\rm Tr} \left( W^{\b}\ast W_{\b} \right) - \pa^2{\rm Tr} \left( \Gamma^{\a} \ast W^{\b} \right)  {\rm Tr} \left(\Gamma_{\a}\ast  W_{\b}\right) \nonumber \\
&& \qquad \qquad \qquad \qquad -{\rm Tr} \left( \pa^2 \Gamma^{\a}\right)  {\rm Tr} \left( \Gamma_{\a} \ast W^{\b} \ast W_{\b} \right) -{\rm Tr} \left( \pa^2 W^{\a} \right)  {\rm Tr} \left( W_{\a} \ast \Gamma^{\b}\ast \Gamma_{\b}\right) \Big]_{\overline{\theta}=0} \nonumber \\
\eea
In this case it would be possible to replace all the star products with ordinary products in the first two terms, but not in the last two.\\
$\bullet$ Antichiral sector:\\
We obtain
\bea
\overline{{\cal D}}^{(4)}&=& - \frac{1}{12}  {\cal S} {\cal F}^2 \int d^8 z~\overline{\theta}^2 \Big[ \pa^2{\rm Tr}\left(\overline{W}^{\adot} \ast \overline{\Gamma}_{\adot} \right) {\rm Tr}\left( \overline{\Gamma}^{\g \gdot} \ast \overline{\Gamma}_{\g \gdot}\right) \nonumber \\
&& \qquad \qquad \qquad \qquad\qquad \qquad \qquad + 8 ~\pa^2{\rm Tr}\left(\overline{\Gamma}^{\adot} \ast \overline{W}^{\bdot} \right) {\rm Tr}\left(  \overline{W}_{\bdot}\ast \overline{\Gamma}_{\adot} \right) \Big] \nonumber \\
&& = - \frac{1}{2}{\cal S} {\cal F}^2\int d^4 x \Big[{\rm Tr}\left(\overline{W}^{\adot}\overline{W}_{\adot}\right){\rm Tr}\left(\overline{W}^{\bdot}\overline{W}_{\bdot}\right)\Big]_{\theta=\overline{\theta}=0}
\eea

\section{(Super)gauge invariance}

Collecting all the results of the previous section and performing the $\theta$ and $\overline{\theta}$ integrations for simplicity, the divergent part of the one--loop gauge effective action reads
\bea
\G^{(1)}_{gauge} &=&  \frac{1}{2}(-3 + N_f) {\cal S} \int d^4x \Bigg\{  \frac{1}{2} D^2 ~\Big[~{\cal N}~{\rm Tr}\left(W^{\a} ~W_{\a}\right)   - {\rm Tr}\left( W^{\a}  \right){\rm Tr}\left(  W_{\a}\right)\Big] \nonumber \\
&& \qquad  + \frac{1}{2} \overline{D}^2 ~\Big[~{\cal N}~{\rm Tr}\left(\overline{W}^{\adot} ~\overline{W}_{\adot}\right)   - {\rm Tr}\left( \overline{W}^{\adot}  \right){\rm Tr}\left(  \overline{W}_{\adot}\right)\Big] \nonumber \\
&& \qquad + i {\cal F}^{\rho\g} D^2~ [ 2{\rm Tr}(D_\rho W^\a)
{\rm Tr}(\G_\g   W_\a)- {\rm Tr}(D_\rho \G_\g) 
{\rm Tr}(W^\a W_\a) ] \nonumber \\
&& \qquad+ i {\cal F}^{\rho \g} \Big[~2~\pa_{\rho \dot{\rho}}{\rm Tr}\left( \overline{W}^{\adot} \right) {\rm Tr}\left(\overline{W}_{\adot} \overline{\Gamma}_{\g}^{~\dot{\rho}} \right) + \pa_{\rho \dot{\rho}}{\rm Tr}\left( \overline{\Gamma}_{\g}^{~\dot{\rho}} \right) {\rm Tr}\left(\overline{W}^{\adot}\overline{W}_{\adot}\right)\Big] \nonumber \\
&& \qquad + {\cal F}^2 ~\Big[ \frac{1}{2} D^2{\rm Tr} \left( \Gamma^{\a} \Gamma_{\a} \right)  D^2 {\rm Tr} \left( W^{\b} W_{\b} \right) - D^2{\rm Tr} \left( \Gamma^{\a} W^{\b} \right)  D^2 {\rm Tr} \left(\Gamma_{\a} W_{\b}\right) \nonumber \\
&& \qquad \qquad -{\rm Tr} \left( D^2 \Gamma^{\a}\right)  D^2 {\rm Tr} \left( \Gamma_{\a} \ast W^{\b} \ast W_{\b} \right) - {\rm Tr} \left( D^2 W^{\a} \right) D^2 {\rm Tr} \left( W_{\a} \ast \Gamma^{\b}\ast \Gamma_{\b}\right) \Big] \nonumber \\
&& \qquad - \frac{1}{2}{\cal F}^2 \Big[{\rm Tr}\left(\overline{W}^{\adot}\overline{W}_{\adot}\right){\rm Tr}\left(\overline{W}^{\bdot}\overline{W}_{\bdot}\right)\Big]\Bigg\}_{\theta=\overline{\theta}=0} \nonumber \\
&& \equiv \frac{1}{2}(-3 + N_f) \big[ {\cal D}^{(2)} +  \overline{{\cal D}}^{(2)} + {\cal D}^{(3)} + \overline{{\cal D}}^{(3)} + {\cal D}^{(4)}+ \overline{{\cal D}}^{(4)} \big]
\label{final}
\eea
where factor $\frac{1}{2}$ comes from (\ref{coef1/2}), $(-3)$ is the contribution from the ghosts whereas $N_f$ comes from matter. We note that the contributions to the two-point functions are independent of ${\cal F}$, three-point functions are linear in ${\cal F}$ and four-point functions are quadratic in ${\cal F}$.\\ 
We consider the variation of $\G^{(1)}_{gauge}$ under supergauge transformation $e_\ast^{V'} 
= e_\ast^{i \overline{\Lambda}} \ast e_\ast^{V}\ast e_\ast^{-i \Lambda}$. 
Superfield strengths and superconnections transform as 
\bea
&& ~~\d \G_\a = D_\a \L + i [ \L , \G_\a]_{\ast} \qquad \qquad \qquad \d W_\a = i [ \L, W_\a ]_{\ast} \nonumber \\
&& \delta \overline{\Gamma}_{\bdot} = \overline{D}_{\bdot}\overline{\Lambda} + i [\overline{\Lambda}, \overline{\Gamma}_{\bdot}]_{\ast} \qquad ~~\delta \overline{\Gamma}_{\b \bdot} = \pa_{\b \bdot}\overline{\Lambda} + i [\overline{\Lambda}, \overline{\Gamma}_{\b \bdot}]_{\ast} \qquad ~~ \delta \overline{W}_{\bdot} = i [\overline{\Lambda}, \overline{W}_{\bdot}]_{\ast} \nonumber \\
\eea
from which we can easily infer the transformation rules of the components appearing 
in (\ref{final}). By expanding the $\ast$-product, after a long but straightforward calculation, it is possible to show that 
\beq
\d {\cal D}^{(2)} = A ~{\cal S}\qquad \qquad \d {\cal D}^{(3)} = - \left( A + B \right)~{\cal S} \qquad \qquad \d {\cal D}^{(4)} = B ~{\cal S}
\eeq
with 
\bea
A &=& 2 i {\cal F}^{\rho \g} \int d^4x \Big[ {\rm Tr} \left ( D_{\g} \Lambda D_{\rho} W^{\a} \right) {\rm Tr} \left ( D^2 W_{\a}\right) -  {\rm Tr} \left ( D^2 \Lambda D_{\rho} W^{\a} \right) {\rm Tr} \left ( D_{\g} W_{a}\right) \nonumber \\
&& \qquad \qquad \qquad \qquad \qquad \qquad\qquad ~ +  {\rm Tr} \left ( D_{\g} \Lambda D^2 W^{\a} \right) {\rm Tr} \left ( D_{\rho} W_{\a}\right) \Big] _{\theta=\overline{\theta}=0} \\ 
B &=& - {\cal F}^2 \int d^4x \Big[ 2 {\rm Tr} \left( D^2 \Lambda D_{\b} W^{\a}\right) D^2 {\rm Tr} \left( \Gamma^{\b} W_{\a} \right) 
-2 {\rm Tr} \left(  D_{\b}\Lambda D^2 W^{\a} \right)D^2 {\rm Tr} \left( \Gamma^{\b} W_{\a}\right) \nonumber \\
&& \qquad \qquad - {\rm Tr} \left(  D^2 \Lambda D_{\b} \Gamma^{\b}\right)D^2 {\rm Tr} \left( W^{\a}W_{\a}\right) 
+ {\rm Tr} \left(  D_{\b}\Lambda D^2 \Gamma^{\b} \right)D^2 {\rm Tr} \left( W^{\a}W_{\a}\right) \nonumber \\
&& \qquad \qquad+ {\rm Tr} \left( D^2 W^{\a}\right){\rm Tr} \left( \left\{ D^{\b} \Gamma^{\g}, D_{\g} \Lambda \right\} D_{\b} W_{\a} \right)
- {\rm Tr} \left( D^2 W^{\a}\right){\rm Tr} \left( \Gamma^{\g} \left\{ D^2 \Lambda , D_{\g} W_{\a} \right\}\right) \nonumber \\
&&\qquad \qquad + {\rm Tr} \left( D^2 W^{\a}\right){\rm Tr} \left( \Gamma^{\g} \left[  D_{\g} \Lambda , D^2 W_{\a} \right]\right) 
+ {\rm Tr} \left( D^2 W^{\a}\right){\rm Tr} \left( \left\{D^2 \Lambda, D_{\g} \Gamma^{\g}  \right\} W_{\a} \right) \nonumber \\
&& \qquad \qquad + {\rm Tr} \left( D^2 W^{\a}\right){\rm Tr} \left( \left[ D^2 \Gamma^{\g}, D_{\g}\Lambda \right]W_{\a} \right) 
+ {\rm Tr} \left( D^2 \Gamma^{\g}\right){\rm Tr} \left( D_{\g} \Lambda D^2 W^{\a} W_{\a}\right) \nonumber \\
&& \qquad \qquad + {\rm Tr} \left( D^2 \Gamma^{\g}\right){\rm Tr} \left( D_{\g} \Lambda W^{\a} D^2 W_{\a} \right) 
+{\rm Tr} \left( D^2 \Gamma^{\g}\right){\rm Tr} \left( D_{\g} \Lambda D_{\b} W^{\a} D^{\b} W_{\a}\right) \nonumber \\
&&\qquad \qquad - {\rm Tr} \left( D^2 \Gamma^{\g}\right){\rm Tr} \left( D^2 \Lambda D_{\g} W^{\a} W_{\a}\right) 
+ {\rm Tr} \left( D^2 \Gamma^{\g}\right){\rm Tr} \left( D^2 \Lambda W^{\a} D_{\g} W_{\a}\right)  \Big]_{\theta=\overline{\theta}=0} \nonumber \\
&& - {\cal F}^2  {\cal F}^{\rho \g} \int d^4x \Big[ ~~{\rm Tr} \left( D^2 W^{\a} \right){\rm Tr} \left( D_{\rho} \Gamma_{\g} \left[ D^2\Lambda, D^2 W_{\a}\right]\right) \nonumber \\
&& \qquad \qquad + {\rm Tr} \left( D^2 W^{\a} \right){\rm Tr} \left( \left[ D^2 \Gamma_{\g} , D^2 \Lambda \right] D_{\rho} W_{\a} \right) 
+ {\rm Tr} \left( D^2 \Gamma_{\rho} \right){\rm Tr} \left( D^2 \Lambda D^2 W^{\a} D_{\g} W_{\a}\right) \nonumber \\
&& \qquad \qquad + {\rm Tr} \left( D^2 \Gamma_{\rho} \right){\rm Tr} \left( D^2 \Lambda D_{\g} W^{\a} D^2 W_{\a} \right) \Big]_{\theta=\overline{\theta}=0} 
\eea
whereas
\bea
&& \d \overline{{\cal D}}^{(2)} = - \d \overline{{\cal D}}^{(3)} \nonumber \\
&& ~\qquad = - 2 i  {\cal S}{\cal F}^{\rho \g} \int d^4x ~ {\rm Tr}\left( \pa_{\rho \dot{\rho}} \overline{W}^{\adot} \right) ~ {\rm Tr}\left( \pa_{\g}^{~\dot{\rho}}\overline{\Lambda}~ \overline{W}_{\adot} \right) \Bigg|_{\theta=\overline{\theta}=0} \\
&& \d \overline{{\cal D}}^{(4)} = 0
\eea 
We note that in the chiral sector the gauge variation, when evaluated in components, is proportional to $D_{\g} \Lambda|$ and $D^2 \Lambda|$ but not to $\Lambda|$. Therefore, in this sector ordinary gauge invariance is preserved term by term, whereas the supergauge one seems to be broken. In the antichiral sector instead, the gauge variation of each term is proportional to $\overline{\Lambda}|$ so breaking ordinary gauge invariance. However it is easy to see that all the variations sum up to zero and we find   
\beq
\d \G^{(1)}_{gauge} = 0
\eeq
We have then proved the supergauge invariance of the one--loop gauge effective
action.


\section{Conclusions}

In this paper we have computed one--loop divergent contributions to
the gauge effective action for N=1/2 $U({\cal N})$ SYM theory
with matter in the adjoint representation of the gauge group.
It turns out that new divergent terms proportional to the NAC 
parameter appear for the three and four--point functions of
the gauge field. Term by term these corrections are
not (super)gauge invariant being proportional not only to 
superfield--strengths but also to superconnections. However, we have proved that
the complete divergent part of the effective action at one--loop 
is {\em supergauge invariant} since nontrivial cancellations occur
among the gauge variations of the two and three point functions 
and of the three and four point functions. This allows us to 
conclude that at least at one--loop there must be super Ward identities
at work which guarantee the invariance of the theory,
despite the appearance of a NAC product which breaks explicitly 
supersymmetry and is not invariant under supergauge transformations.
Therefore, a pattern similar to the one studied for SYM theories
with bosonic noncommutativity \cite{PSZ} seems to be present also in this case.

The analysis has been carried on by using a manifestly gauge invariant
superspace setup. This has been accomplished by working directly with
the star product (no expansions at the level of the action), performing
Fourier transform also on the spinorial coordinates so trading the star
products for spinorial phases and adapting the $D$--algebra in order
to reduce the supergraphs to ordinary momentum integrals.   

In order to study the supergauge invariance of the result we have
found convenient to generalize the background field method to the
NAC theory. In this approach the contributions to the effective
action turn out to be proportional to the geometric objects of the
theory, i.e. superconnections and superfield--strengths. 
The generalization of the background field method to the NAC case has revealed not straightforward for two main reasons: The hermitian conjugation rules in the classical action change in euclidean signature and some important identities involving covariant derivatives do not hold anymore because of the presence of a noncommutative product. 
   
Our approach is particularly useful when the relation between
supergauge invariance and NAC geometry at higher loops is of concern.
Moreover, it makes possible the evaluation of higher--loop corrections
to the effective action and higher--loop correlation
functions for composite operators which in general is quite prohibitive
in components. In principle, our method could be also generalized to the 
study of SYM theories with extended supersymmetry not broken \cite{FS} or
partially broken by nonanticommutativity \cite{ivanov}.

In this paper we have focused on the supergauge invariance of the
gauge part of the effective action. Divergent contributions with matter on the external lines have still to be computed. Moreover, 
renormalizability of NAC SYM theories in superspace and the 
relation of our results with the ones found in components \cite{AGS, JONES}
have not been discussed yet. This all will be the subject of a future publication 
\cite{GPR2}.

\vskip 30pt
{\bf Acknowledgments} We thank M.T. Grisaru for his fundamental contribution
at the early stage of the work. This work has been supported in
part by INFN, COFIN prot.  2003023852\_008 and the European Commission RTN 
program MRTN--CT--2004--005104.

\end{document}